\documentclass[journal=jacsat,manuscript=article]{achemso}

\usepackage[version=3]{mhchem} 
\usepackage{array}
\usepackage{hyperref}
\usepackage{xcolor, soul}

\author{Laku Dorjee Tamang}
\affiliation{Advanced Functional Materials \& Simulation Lab (AFMSL), Department of Physics, Mizoram University, Aizawl,	796004, India}
\alsoaffiliation{Physical Sciences Research Center (PSRC), Department of Physics, Pachhunga University College, Aizawl, 796001, India}
\author{Shivraj Gurung}
\affiliation{Physical Sciences Research Center (PSRC), Department of Physics, Pachhunga University College, Aizawl, 796001, India}
\author{Bhanu Chettri}
\affiliation{Advanced Functional Materials \& Simulation Lab (AFMSL), Department of Physics, Mizoram University, Aizawl,	796004, India}
\author{Nguyen Thanh Tien}
\affiliation{College of Natural Sciences, Can Tho University, 
Can Tho City, 900000, Viet Nam}
\author{Le Huu Nghia}
\affiliation{College of Natural Sciences, Can Tho University, 
Can Tho City, 900000, Viet Nam}
\author{Darwin Barayang Putungan}
\affiliation{Institute of Physics, College of Arts and Sciences, University of the Philippines Los Banos, Laguna, 4031, Philippines}

\author{Ranjit Thapa}
\affiliation{Department of Physics, SRM University AP, Amaravati, Andhra Pradesh 522 240, India}
\alsoaffiliation{Centre for Computational and Integrative Sciences, SRM University AP, Amaravati, Andhra Pradesh 522 240, India}
\email{ranjit.phy@gmail.com}

\author{Kailash Chandra Bhamu}
\email{kcbhamu85@gmail.com}
\affiliation{Department of Physics, SLAS, Mody University of Science and Technology, Lakshmangarh, Sikar, Rajasthan, 332311, India}

\author{Dibya Prakash Rai}
\affiliation{Advanced Functional Materials \& Simulation Lab (AFMSL), Department of Physics, Mizoram University, Aizawl,	796004, India}
\email{dibyaprakashrai@gmail.com}

\title[An \textsf{achemso} demo]
  {A comparative first-principles investigation of bilayer NbOX$_2$ (X=Cl, Br, I) for Photocatalytic water splitting applications.}

\abbreviations{IR,NMR,UV}
\keywords{DFT; GGA-PBE; HSE06; Photocatalysis; HER; Absorption coefficient}

\begin{document}

\begin{tocentry}

    \includegraphics[width=0.98\linewidth]{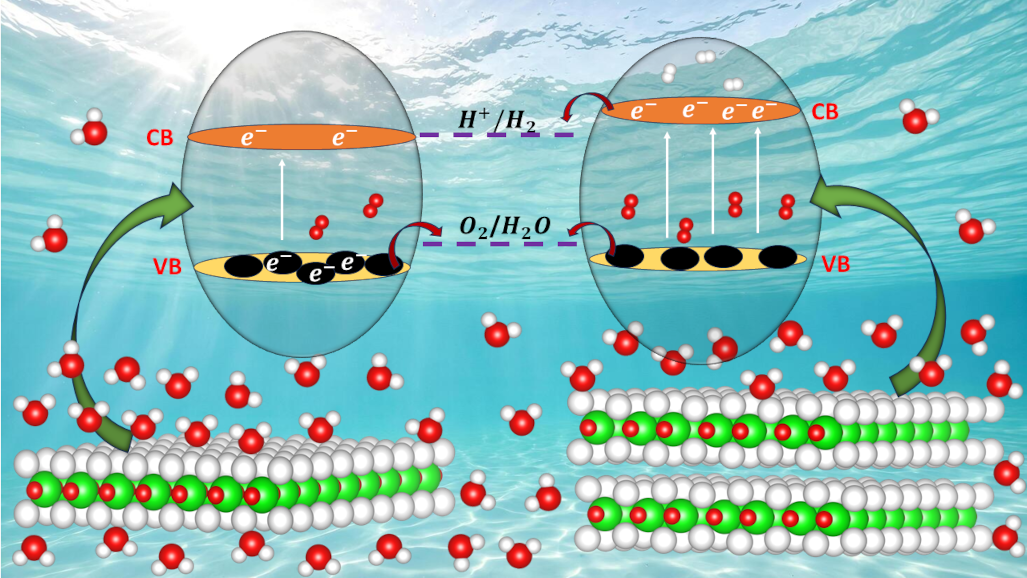}
 
\end{tocentry}

\begin{abstract}
Motivated by our previous work on bulk NbOX$_2$, where we have reported its high dielectric polarisation and finite piezoelectric response, this work extends to its 2D homo bilayer system to explore its potential for photocatalytic water splitting. Herein, density functional theory (DFT) were employed in probing the structural, electronic, optical, and photocatalytic properties of 2D homo bilayer NbOX$_2$ (X = Cl, Br, and I). Our results show that structurally, NbOCl$_2$ and NbOBr$_2$ prefer AC bilayer stacking, while AB stacking was preferred by NbOI$_2$. All the considered bilayers are dynamically, thermally, and mechanically stable. From the analysis of electronic structure we have found a decreasing trend in the energy band gap as X goes down the group from Cl to I, with the position of the valence band maximum shifting upward along the high symmetry points. In terms of carrier mobility, all 2D bilayer systems possess high carrier mobility comparable to known 2D materials. It also exhibits an anisotropic carrier transfer property by which charge carriers are separated efficiently. These materials show similar trends to BiOI and PtSe$_2$, in which photocatalytic efficiency was increased by forming the multiple layers. The materials under investigation are suitable for photocatalytic water splitting under visible and ultraviolet regions with absorption coefficients of (10$^5$ cm$^{-1}$).
\end{abstract}

\section{Introduction}
In today's world, energy is crucial for economic development and the sustainability of life. However, the rapid depletion of energy resources is a greater threat. For centuries, fossil fuels have played a major role in human civilisation and industrial development. Meanwhile, fossil fuel reservoirs have been degrading day by day due to overutilization, leading to the emission of harmful gases that contribute to environmental pollution. Now, people have realised its credibility and have become more cautious about energy consumption and its adverse effect on the health. The governments are more focused on solving this issue and are initiating various energy-oriented research programs. The scientists are looking for innovative alternatives to harvest green and clean energy without compromising its efficacy, balancing sustainability and development. The primary focus is on the renewable sources of energy, such as solar cells\cite{al2022photovoltaic}, hydropower\cite{yuksel2010hydropower}, geothermal\cite{barbier2002geothermal}, wind power\cite{csahin2004progress}, piezoelectric\cite{tamang2025electronic,tressler1998piezoelectric}, and photocatalysis\cite{maeda2010photocatalytic,wang20182d,subba20252d,subba_cl}, etc. Among the energy resources, Hydrogen is a clean form of energy, very light in weight, high energy density (122 kJ mol$^ {-1}$) and highly combustible, giving the carbon-free byproduct as H$_2$O.  However, generating H$_2$ and storing it in compressed form is a challenging task\cite{albuquerque2024advances,subba2026insights}.
Photocatalysis is one such method to produce H$_2$ from H$_2$O molecules (2H$_2$O $\rightarrow$ 2H$_2$+O$_2$) by trapping the sunlight with the help of a solid-state semiconductor electrode\cite{laku2025}.
In 1972, Fujishima and Honda proposed a novel, cost-effective and eco-friendly method (photocatalysis) to split the water into hydrogen and oxygen by using a semiconductor electrode and solar energy \cite{fujishima1972electrochemical}. Moreover, identifying a suitable semiconductor electrode for effective photocatalysis is challenging. Several materials been explored both theoretically and experimentally and identified as an efficient photocatalyst based on the following criteria: (i) they must possess a band energy $\ge$ 1.23 eV, (ii) Proper band edge position with relative to redox potential of water (i.e, E$_H^+/H_2$ = -4.44 eV, E$_{O_2}/H_2O$ = -5.67 eV)\cite{ji2018janus} (iii) a strong light harvesting ability\cite{pan2022two} and (iv) high carrier mobility of photo-excited electrons and holes at the water interface\cite{cao2015polymeric}.
The photocatalysis technique has promises diverse applications such as in environmental remediation, including dye degradation,\cite{huang2017coupling} water purification, the treatment of domestic sewage, the removal of harmful indoor gases,\cite{dalton2002photocatalytic,huang2020synthesis} the production of solar cells,\cite{zhang2013dye,bagher2015types} water splitting for H$_2$, O$_2$, and H$_2$O$_2$,\cite{tian2017precursor,stern2015ni} and carbon dioxide reduction and so on.\cite{chalk2006key,peschka2012liquid,omer2008energy} However, the benchmark efficiency has not been achieved for real-time application due to various deficiencies like limited charge separation lead to high rate of photocorossion, low active sites at surface, low electrical conductivity, surface corrosion, structural instability in water, mismatch of band-edge with redox potentials, low illumination site limit light absorption, and lack of tunability like nanoparticles. In general, the photocatalytic process consists of three stages: 
\begin{enumerate}
    \item Induction of photoexcited electron and hole pairs by irradiation of light (h$\nu \ge E_g$).
    \item 
    diffusion of electron and hole pairs to the active sites of photocatalysts, during which some of them will be recombined.
    \item The remaining electrons and holes will be used for a redox reaction.
\end{enumerate}

Following the successful derivation of two-dimensional (2D) graphene in 2004, 2D materials have become the centre of research in a new era for energy harvesting and nano-devices, due to their interesting physical, chemical, optical, and electronic properties\cite{novoselov2004electric}. Various families of 2D materials have been investigated both experimentally and theoretically to study their intrinsic properties, such as molybdenum diselenide (MOSe$_2$)\cite{bromley1972band}, tungsten disulfide (WS$_2$)\cite{houben2012diffraction}, black phosphorus (BP)\cite{liu2014phosphorene}, ferric oxide (Fe$_2$O$_3$)\cite{chen2014nanostructure}, transition metal dichalcogenides(TMD)\cite{wang2013electrochemical,xu2014spin}, metal-organic frameworks(MOFs)\cite{zhao2015ultrathin,clough2015two}, covalent organic frameworks (COFs)\cite{colson2011oriented,spitler2010lewis}, perovskites and many more. Nonetheless, these materials exhibited strong potential for a large extent of chemical and physical applications; however, they have some limitations due to intrinsic defects\cite{chen2017widely,zhao2017high}. Therefore, substantial efforts have been made in exploring the properties of 2D materials. A heterojunction of graphitic carbon nitride (g-C$_3$N$_4$) and Bi$_2$O$_2$Se  strategy has been adopted to address the issue of utilisation of low efficiency of photogenerated charges in nanosheets \cite{lin2024improving}. \par
In recent years, numerous works have been done both theoretically and experimentally on niobium oxide dihalides NbOX$_2$ (X = Cl, Br, and I) for energy harvesting, which belongs to a group of family transition metal oxide dihalides MOX$_2$ (M= V, Nb, Mo, and Ta, and  X = Cl, Br, and I)\cite{schnering1963aufbau,hillebrecht1997structural}. Since the 1960s, bulk NbOX$_2$ have been synthesized, and as van der Waals-layered materials, their corresponding monolayers have been obtained via mechanical exfoliation techniques. Moreover, their structure is similar to that of a perovskite crystal ABO$_3$, in which each monolayer NbOX$_2$ consists of NbO$_2$X$_4$ octahedra where X-X and O atoms were interconnected along the crystallographic a and b axes, respectively. These monolayers have been studied in the field of piezoelectric, capacitor, and photocatalytic applications due to their intriguing electronic, structural and optoelectronic properties. A previous study reported that NbOX$_2$ monolayers exhibit an indirect band gap within the range of (1.77-1.88)eV\cite{jia2019niobium} with a good light harvesting ability, and their solar-to-hydrogen efficiency at different PH levels was also reported\cite{wan2025first}. 
Among these monolayers, NbOI$_2$ has been synthesised experimentally using a chemical vapour transport method \cite{fang20212d} and the exfoliation energies of NbOCl$_2$ and NbOBr$_2$ were predicted to be lower than those of graphene, indicating their feasibility in the near future\cite{wang2015measurement}.\par
The objective of this paper is to study the stacking effects on the structural, electronic, optical and photocatalytic properties of homo-bilayer NbOX$_2$ (X= Cl, Br, and I) using a first principle approach since numerous studies have reported that by forming a homo-bilayer, one can greatly enhance the material properties, such as its stability, conductivity, light harvesting ability and photocatalytic properties\cite{deb2022bilayer,wei2019investigation,zhang2019stacking}.

\section{Computational Details}

All calculations were performed using density functional theory (DFT), incorporated in the Vienna ab initio simulation package (VASP)\cite{kresse1996efficient}. The projector augmented-wave (PAW) potentials were used to characterise the core electrons\cite{blochl1994projector,kresse1999ultrasoft}. The Perdew-Burke-Ernzerhof scheme within the generalised gradient approximation (GGA-PBE) was used to describe the exchange-correlation interaction among electrons\cite{perdew1996generalized}. Since the PBE-GGA functional underestimates the van der Waals (vdW) interactions of bilayer structures, Grimme's DFT-D2 correction was also included \cite{grimme2010consistent}. A plane wave of kinetic energy with a cutoff of 540 eV and a k-mesh of 8 $\times$ 8 $\times$ 1 was adopted for structural optimisation and a Monkhorst pack grid of 12 $\times$ 24 $\times$ 1 for SCF calculation. To avoid interactions between the neighbouring interlayer, a large vacuum of 30 \AA was added along the z-axis. The convergence criteria for ionic forces and energy were set to $10^{-3}$ eV \AA$^{-1}$ and $10^{-7}$ eV, respectively. As an accurate prediction of the electronic band gap and optical properties is essential for understanding the photocatalytic performance of a material, and GGA-PBE, LDA and some other approximations are known to underestimate the electronic band gap. Hybrid functionals such as HSE06, B3LYP, and GW are widely adopted for correctly predicting the electronic band gap. Hence, in this study, the electronic structure and optical properties are calculated using the HSE06 hybrid functional with a plane wave having a cut off of 520 eV and a K-mesh density of $12\times 24 \times 1$.

\section{Results and discussion}
\subsubsection{Structural Properties, Dynamical And Mechanical Stability}
In this study, the monolayer NbOX$_2$ (X = Cl, Br, and I) was adopted from \cite{mortazavi2022highly}, which belongs to the space group P1. Relaxed structural parameters obtained by the PBE method are summarized in Table1, which are in agreement with previously reported works\cite{mortazavi2022highly,ye2023manipulation,helmer2025computational}. Then we studied the layer-stacking behaviour of NbOX$_2$. Three possible stacking configurations were considered for each monolayer, namely AA, AB and AC as shown in Fig.\ref{1}. For the AA stacking configuration, the upper layer is perfectly placed with the under layer in the xy-plane. In the case of AB stacking, the upper layer is shifted along the a-direction such that the Nb atom is on top of the halogen atom (as shown in Fig.\ref{1}(f)). For the AC stacking, the upper layer is shifted in the b-axis such that the Nb atom is on top of the O atom (as shown in Fig.\ref{1}(d)). All the relaxed structures of the homo-bilayer exhibit triclinic geometry with a space group P1, which is similar to their corresponding monolayer. To check the stability of the different stacking configurations, after geometry optimisation, we calculated their ground state energy and compared it, which is summarised in Table(s1). Among different stacking configurations of the bilayer NbOX$_2$ (X= Cl, Br, and I), NbOCl$_2$ and NbOBr$_2$ with AC stacking and NbOI$_2$ with AB stacking were the most stable configurations according to their ground state energy. Hence, they were taken for further calculations. The relaxed structure and their respective lattice parameters under investigation were given in Table \ref{t4}, which are in consistent with the theoretical value obtained previously\cite{mortazavi2022highly}. The interlayer distances of the considered bilayer NbOCl$_2$, NbOBr$_2$, and NbOI$_2$ were 2.887 $\AA$, 3.159 $\AA$, and 3.200 $\AA$, respectively.
The optimised lattice parameters show a strong anisotropy due to different compositions along the c (X-Nb-X) and b (O-Nb-O) directions. In these crystals, due to spontaneous polarisation, the cation Nb is displaced from its central position, resulting in two different Nb-O and Nb-X distances.\cite{whangbo1982importance,rijnsdorp1978crystal} Moreover, the bond lengths and lattice constant become larger as we move down the group from Cl to I because of the increase in their respective atomic radii. 
Here, we have focused primarily on NbOX$_2$ bilayer properties rather than monolayers, since monolayers have been studied thoroughly by the Pan {\it et al.} group\cite {pan2023two,mortazavi2022highly}.
\begin{table}[h]
    \centering
     \begin{tabular}{|p{2.00cm}|c|c|c|c|c|c|}
    \hline
    NbOX$_2$ & \multicolumn{3}{|c}{Lattice Constants($\AA$)}  & \multicolumn{3}{|c}{Angle ($\deg$)} \\
    \hline
    & a & b & c & $\alpha$ & $\beta$ & $\gamma$ \\
    Bilayer & & & & & & \\
    X = Cl & 6.699 & 3.901 & 33.702 & 89.74 & 90.04 & 90.00  \\
    X = Br & 7.024 & 3.898 & 34.908 & 89.84 & 90.04 & 90.01  \\
    X = I & 7.534 & 3.908 & 36.41 & 90.00 & 93.11 & 90.00 \\
    \hline   
    Monolayer &  &  &  & & &  \\
    X = Cl & 6.751 & 3.928 & 30.00 & 90.00 & 90.00 & 90.00 \\
    X = Br & 7.079 & 3.930 & 20.00 & 90.00 & 89.99 & 90.00 \\
    X = I & 7.585 & 3.939 & 32.00  & 89.99 & 89.99 & 90.71 \\
    \hline
    \end{tabular}
    \caption{Obtained values of lattice parameter of monolayers and bilayer NbOX$_2$(X = Cl, Br, and I), optimised by PBE-GGA functional.}
    \label{t4}
\end{table}
We evaluated the thermodynamic stability of bilayer NbOX$_2$ by calculating {\it ab initio} molecular dynamics (AIMD) simulations performed at 300 K with a time step of 2 fs
under the NVT canonical ensemble. As seen in Fig.\ref{2}, no major distortion or broken bonds were present after the 6ps simulation, indicating their thermal stability at room temperature. We also check their experimental feasibility by calculating the cohesive energy using an equation (\ref{eq1})
\begin{equation}
    E_{coh} = \frac{N_{Nb}E_{Nb} + N_OE_O + N_XE_X - E_{BL}}{N_{tot}}
    \label{eq1}
\end{equation}
Where, N$_{Nb}$, N$_O$, and N$_X$ are the numbers of Nb, O, and X atoms per unit cell, respectively. N$_{tot}$ is the total number of atoms in each bilayer. E$_{Nb}$, E$_O$, E$_X$, and E$_{BL}$ are energies of isolated Nb, O, X atoms and their total ground state energy of the composed bilayer, respectively.
It is reported that a positive E$_{coh}$ indicates bond formation among atoms within the material; thus, the bigger the value, the more stable the structure. The obtained values of E$_{coh}$ are 6.45 eVatom$^{-1}$, 6.15 eVatom$^{-1}$, and 5.83 eVatom$^{-1}$ for NbOCl$_2$, NbOBr$_2$, and NbOI$_2$, respectively, which are similar to those already synthesised and theoretically predicted\cite{frank2019many,sun2021ultrahigh}, implying their experimental possibility in future. As shown in Fig.(S1), the phonon dispersion spectrum indicates that these bilayers are dynamically stable even though there is a negligible amount of negative frequency in the phonon spectra.\cite{sun2020b2p6}\par
 
\begin{figure}[H]
		\centering
		\includegraphics[width=1.0\textwidth]{bilayer.png}
		\caption{Schematic representation of monolayers(a-c) and its respective bilayer(d-f) with different stacking configuration.}
		\label{1}
	\end{figure} 

\begin{figure}[H]
		\centering
		\includegraphics[width=1.1\textwidth]{AIMD.png}
		\caption{Snapshots of AIMD calculations of bilayer NbOX$_2$ at 300K.}
		\label{2}
	\end{figure} 

\begin{table*}[htp]
    \centering
     \begin{tabular}{p{2.0cm} p{1.35cm} c c c c c c c c}    
      
     Systems &C$_{11}$  & C$_{12}$  & C$_{22}$  & C$_{66}$ & E (Nm$^{-1}$) & G  & K & $\nu$ & $\kappa$ \\
     \hline
     Monolayer\cite{pan2023two} & & & & & & & & &  \\
      X = Cl & 65.16 & 5.37 & 92.67 & 15.26 & - & - & - & - & -  \\
      X = Br & 63.21 & 4.82 & 88.74 & 14.54 & - & - & - & - & - \\
      X = I & 62.24 & 4.46 & 73.68 & 13.75 & - & - & - & - & - \\
      
      monolayer$^*$
      X = Cl & 66.15 & 5.29 & 93.35 & 15.87 & 32.47 & 13.39 & 18.83 & 0.21 & 1.36 \\
      X = Br & 64.394 & 5.27 & 87.81 & 15.02 & 31.05 & 12.79 & 18.09 & 0.21 & 1.41  \\
      X = I & 62.98 & 4.37 & 79.46 & 14.33 & 29.09 & 12.01 & 16.77 & 0.21 & 1.40 \\

     Bilayer$^*$ & & & & & & & & &  \\
      X = Cl & 131.93 &  11.04 & 203.39 & 31.40 & 202.47 & 49.54 & 87.31 & 0.46 & 1.76  \\
     X = Br & 134.14& 8.85 & 204.16 & 29.63 & 203.58 & 48.82 & 87.09 & 0.49 & 1.78  \\
     X = I & 126.22& 9.539 & 174.39 & 27.571 & 173.66 & 44.15 & 78.89 & 0.48 & 1.79  \\
      \hline
    \end{tabular}
    \caption{The obtained values of elastic constants (C$_{11}$, C$_{12}$, C$_{22}$, and C$_{66}$), young's modulus (E), shear modulus (G), bulk modulus (K) in (unit of Nm$^{-1}$), Poisson's ratio ($\nu$), and Pugh's ratio ($\kappa$) of monolayer and homo-Bilayer NbOX$_2$ (X = Cl, Br, and I) using GGA-PBE functional. (*$\rightarrow$our work)}
    \label{t1}
\end{table*}

\begin{figure}[H]
		\centering
		\includegraphics[width=0.8\textwidth]{ELF.png}
		\caption{Electron localization function of the bilayer (a) NbOCl$_2$, (b)NbOBr$_2$, and (c)NbOI$_2$ respectively.}
		\label{3}
	\end{figure}

Lastly, their mechanical stability was also evaluated by computing their elastic constants C$_{11}$, C$_{12}$, C$_{22}$, and C$_{66}$. To check whether the bilayer NbOX$_2$ is mechanically stable or not, firstly, we calculated the elastic constant of monolayers and found it to be consistent with the value reported previously \cite{pan2023two}. Then, only the elastic constant of bilayers was calculated. The obtained value of the elastic constants is summarised in Table\ref{t1}, and they are found to be mechanically stable by fulfilling the Born\cite{born1996dynamical} stability criteria given in equations\ref{eq3}.
\begin{equation}
    C_{11} > 0, \\
    C_{66} > 0, \\
    C_{11} \times C_{22} > C_{12} \times C_{12}
    \label{eq3}
\end{equation}
From the Table \ref{t1}, it is seen that all the values of bulk moduli are larger than the shear moduli, which implies that the system under investigation offered higher resistance to axial compression than shear deformation. The Pugh's ratio $\kappa$ calculated by B/G of bilayer is given in Table\ref{t1}, and it gives the material's failure mode. When a value of $\kappa$ is below(above) the critical value 1.75, the materials can be considered as brittle(ductile). An alternative method to determine the material's failure is to calculate Poisson's ratio ($\nu$). If the critical value of $\nu$ is below 0.26 (above 0.26), the material is regarded as brittle (ductile). From our calculations, it is observed that bilayer NbOX$_2$ are ductile in nature because of its $\kappa$ and $\nu$ values higher than critical values 1.75 and 0.26. Moreover, the elastic constants of bilayer NbOX$_2$ have also been increased almost by 2 times the value of monolayer NbOX$_2$ as reported by Pan {\it et al.}, which leads to improving their structural stability \cite{pan2023two}.
Hence, the studied bilayers were thermally, dynamically, and mechanically stable.  

\subsection{Electronic Properties}
\begin{figure}[H]
		\centering
		\includegraphics[width=1.0\textwidth]{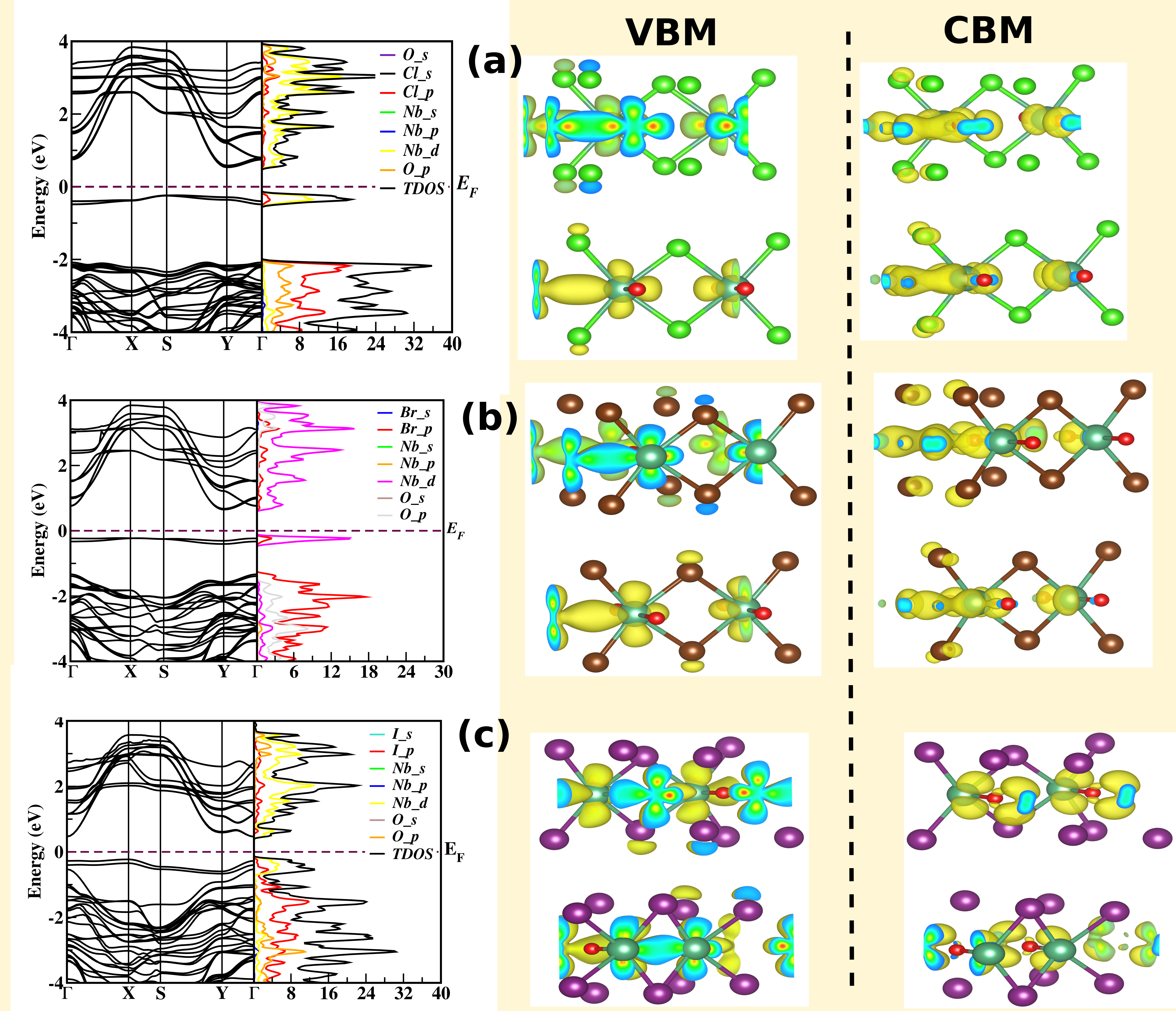}
		\caption{Electronic band structure, density of states, and their corresponding partial charge densities of CBM and VBM (yellow) with the isosurface value of 0.0035 e$\AA$ $^{-3}$ for the bilayers (a) NbOCl$_2$, (b) NbOBr$_2$, and NbOI$_2$, respectively. calculated from PBE-GGA functional.}
		\label{4}
	\end{figure}

To see the effect of stacking on the electronic properties of homo-bilayer NbOX$_2$ (X = Cl, Br, and I), we evaluated their band structure, density of states (DOS), and partial charge density of states (PDOS) (as shown in Fig.\ref{4}) under the HSE06 scheme. The values of band gap obtained from both PBE and HSE06 functional are summarised in Table \ref{t2}. As compared to monolayers, the band gap has been reduced due to the stacking effect. Moreover, it is also observed that the band gap of a homo-bilayer formed by AB and AC stacking has a higher value than that of the naturally occurring bilayer, and here it is approximately 7\%  higher than the value reported earlier\cite{mortazavi2022highly}. It is also observed that stacking does not affect the band nature of the compounds NbOX$_2$, thus preserving its indirect band nature\cite{guo2023ultrathin}, unlike MoS$_2$ where its band nature changes from direct to indirect when a bilayer is formed\cite{conley2013bandgap}. Here, for the calculation of electronic properties, we have used a hybrid functional HSE06 for precise estimation of energy band profile which has always been underestimated by PBE-GGA, especially for those materials exhibiting van der Waals (vdW) interactions \cite{borlido2019large,chen2022electronic}. 
The obtained band values of the respective Monolayers and homo-bilayers are summarised in Table \ref{t2}, which are in agreement with the previously reported values. Further, we studied the contribution of different atomic orbitals to the valence band maximum(VBM) and conduction band minimum(CBM), as shown in Fig.\ref{4}. According to the DOS plot, the p mainly constitutes the lower portion of VBM and d hybridised orbitals. In contrast, the isolated electronic states below the Fermi level are from Nb-d orbitals. which is consistent with the partial charge density plot. Even the CBM is mostly contributed by Nb d orbitals, with a small input from the p orbitals of anion atoms. From the partial charge density plot, it is observed that there is a large electron cloud between the alternative Nb-Nb pairs along the b-direction because of the dimerization effect. The size of an electron cloud between the Nb atoms in NbOI$_2$ is smaller than that in NbOCl$_2$ due to neutralised dimerization effect by large atomic radii\cite{helmer2025computational}. The band structure of monolayers NbOX$_2$ are presented in Fig.S9, which shows similar electronic properties as the homo-bilayer\cite{jia2019niobium,tamang2025electronic,guo2023ultrathin}. 

We also performed the electron localisation function (ELF) to check whether the electrons are localised or delocalised on a particular atom. The ELf plots of bilayer NbOX$_2$, which are represented in Fig.\ref{3}. Where we observed that electrons are mostly localised over the halogen and oxygen atoms. Moreover, the covalent bond changes to an ionic bond down the group due to an increase in size and a decrease in electronegativity. Thus, NbOI$_2$ preferred to form an ionic bond rather than a covalent bond as compared to NbOCl$_2$. 
 
\begin{table}[h]
    \centering
    \begin{tabular}{c c c c c}
    
    NbOX$_2$ & E$_g$(eV)  & E$_{CBM}$(eV) & E$_{VBM}$(eV) & Refs.\\
    \hline
    Bilayer &  &  &  & \\
    X = Cl & 1.74 & - & -  & \cite{mortazavi2022highly}\\
    X = Br & 1.70 & - & - & \cite{mortazavi2022highly} \\
    X = I & 1.62 & - & - & \cite{mortazavi2022highly} \\
    
     Bilayer&  & & & our work \\
   
    X = Cl & 1.86(HSE06) & -4.40 & -6.27 & \\
    X = Br & 1.85(HSE06) & -4.18 & -6.03 & \\
    X = I & 1.74(HSE06) & -3.96 & -5.70 & \\

     Monolayer&  &  &  & \\
    X = Cl & 1.77 & - & - & \cite{mortazavi2022highly} \\
    X = Br & 1.76 & - & - & \cite{mortazavi2022highly} \\
    X = I & 1.70 & - & - & \cite{mortazavi2022highly} \\
    
     Monolayer& & & & our work \\
    X = Cl & 1.93 (HSE06) & -4.45 & -6.38 & \\ 
    & 0.91(GGA) &  & &  \\
    X = Br & 1.91 (HSE06) & -4.49 & -6.40 & \\ 
    & 0.89 (GGA) & - & - & \\
    X= I & 1.84 (HSE06)  & -3.91 & -5.75 & \\
    & 0.86 (GGA)& & & \\
    \hline
    \end{tabular}
    \caption{Calculated values for Band Energy (E$_g$), the potential energies of conduction band minimum (E$_{CBM}$), and valence band maximum (E$_{VBM}$) relative to the vacuum level, respectively for Bilayer as well as monolayer NbOX$_2$.}
    \label{t2}
\end{table}

Further, we also investigated the charge redistribution across the interface of the bilayers(as shown in Fig.\ref{5}) to gain insights into charge transfer between the junction. From the Fig.\ref{4}, we can conclude that there is a transfer of charges between the layers because a large amount of accumulation and depletion regions of charges were seen in the junction.

\begin{figure}[H]
		\centering
		\includegraphics[width=1.0\textwidth]{chg-dn.png}
		\caption{Charge density difference plot of bilayer NbOX$_2$, visualizes in vesta with the iso value of 4.25$\times$ e$^{-05}$ e$\AA$$^{-3}$. In which the yellow and blue region in space represents charge accumulation and depletion, respectively.}
		\label{5}
	\end{figure}
For qualitative analysis of charge transfer between the layers, we calculated the charge density difference by using equation \ref{eq8}.
\begin{equation}
    \Delta \rho = \rho_{(bilayer)} - \rho_{(top layer)} - \rho_{(bottom layer)}
    \label{eq8}
\end{equation}

And for a quantitative study, we performed a Bader charge analysis, from which it is found that 0.001885 $\left|e\right|$, 0.022644 $\left|e\right|$, and 0.00391 $\left|e\right|$ charges had been transferred from the top layer to the bottom layer in NbOCl$_2$, NbOBr$_2$, and NbOI$_2$, respectively. Therefore, the induced electric field at the junction will be beneficial for the photocatalytic properties because it improves the separation of charges by preventing recombination of photo-excited electron-hole pairs. Thus, the catalytic properties of the homo-bilayer will be enhanced as compared to the monolayer NbOX$_2$(X = Cl, Br, and I).

\subsection{Optical properties}
The material's ability to generate electron-hole pairs upon photon absorption determines its potential for photocatalytic water splitting. To evaluate its light-harvesting properties, the Independent Particle Approximation (IPA) method was implemented. 
The response of the materials to the stimuli of the electromagnetic waves mainly depends on their dielectric function $\epsilon(\omega)$, which consists of both real and imaginary parts and can be written as $\epsilon(\omega)$ = $\epsilon_1(\omega)$ + i$\epsilon_2(\omega)$.
where $\epsilon_2$($\omega)$ is related to the band structure given by an equation\ref{eq12}.
\begin{equation}
    \epsilon_2(\omega) = \frac{4\pi e^2}{m^2\omega^2}\sum_{i,f} \int \frac{2d^3k}{(2\pi)^3} |< ik|P|fk >|^2 F_i^k(1-F_j^k)\delta (E_f^k-E_i^k-E)
    \label{eq12}
\end{equation}
Where P, F, E, $\omega$ $\mid$ik$>$, and $\mid$fk$>$ represent the transition matrix, Fermi function, Photon energy, Photon frequency, conduction band states, and valence band states, respectively. The real part $\epsilon_1 (\omega)$ can be derived from Kramers-Kronig relations\cite{kuzmenko2005kramers}.
\begin{equation}
    \epsilon_1(\omega) = 1 + \frac{2}{\pi}P \int _0 ^\infty d\omega ' \frac{(\omega ')\epsilon_2(\omega)}{(\omega')-(\omega)^2}
    \label{eq13}
\end{equation}
where P $\rightarrow$ integral principal value.
The following are the equations that can be used to investigate the optical properties of the materials.
\begin{equation}
   L(\omega) = Im(-\frac{1}{\epsilon(\omega)}) = \frac{\epsilon_2(\omega)}{\epsilon_1 ^2(\omega + \epsilon_2 ^2(\omega)} 
   \label{eq14}
\end{equation}
\begin{equation}
    n (\omega) = \sqrt{\frac{\epsilon_1 ^2(\omega) + \epsilon_2 ^2(\omega) + \epsilon_1(\omega)}{2}}
    \label{eq15}
\end{equation}
\begin{equation}
    \kappa(\omega) = \sqrt{\frac{(\epsilon_1(\omega))^2 + \epsilon_2(\omega)^2)^{1/2} - \epsilon_1(\omega)}{2}}
    \label{eq16}
\end{equation}
\begin{equation}
    \alpha(\omega) = \sqrt{2\omega}[\sqrt{\epsilon ^2(omega) + \epsilon_2 ^2(\omega)} - \epsilon_1(\omega)]^{1/2}
    \label{eq17}
\end{equation}
\begin{equation}
    R(\omega) = \frac{(n-1)^2 + \kappa^2}{(n+1)^2 + \kappa^2}
    \label{eq18}
\end{equation}
Where  L$(\omega)$,  n $(\omega)$, $\kappa$($\omega$), $\alpha$($\omega$), and  R($\omega$) represent the electron energy loss function, refractive index, extinction coefficient, absorption coefficient and the reflectivity of the materials, respectively.

\begin{figure}[H]
		\centering
		\includegraphics[width=0.8\textwidth]{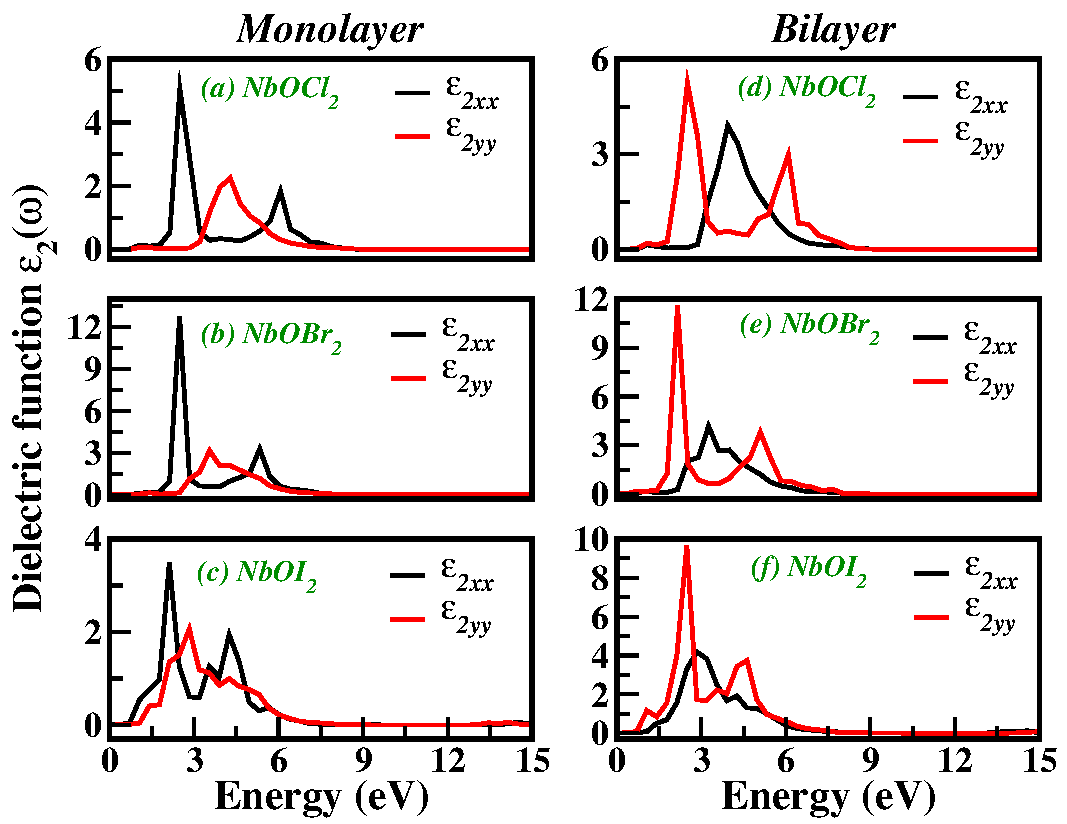}
		\caption{Illustration of the photon absorption by the NbOX$_2$ monolayers(a-c) and bilayers(d-f) in the in-plane (xx and yy) directions, respectively.}
		\label{6}
	\end{figure}

\begin{figure}[H]
		\centering
		\includegraphics[width=0.8\textwidth]{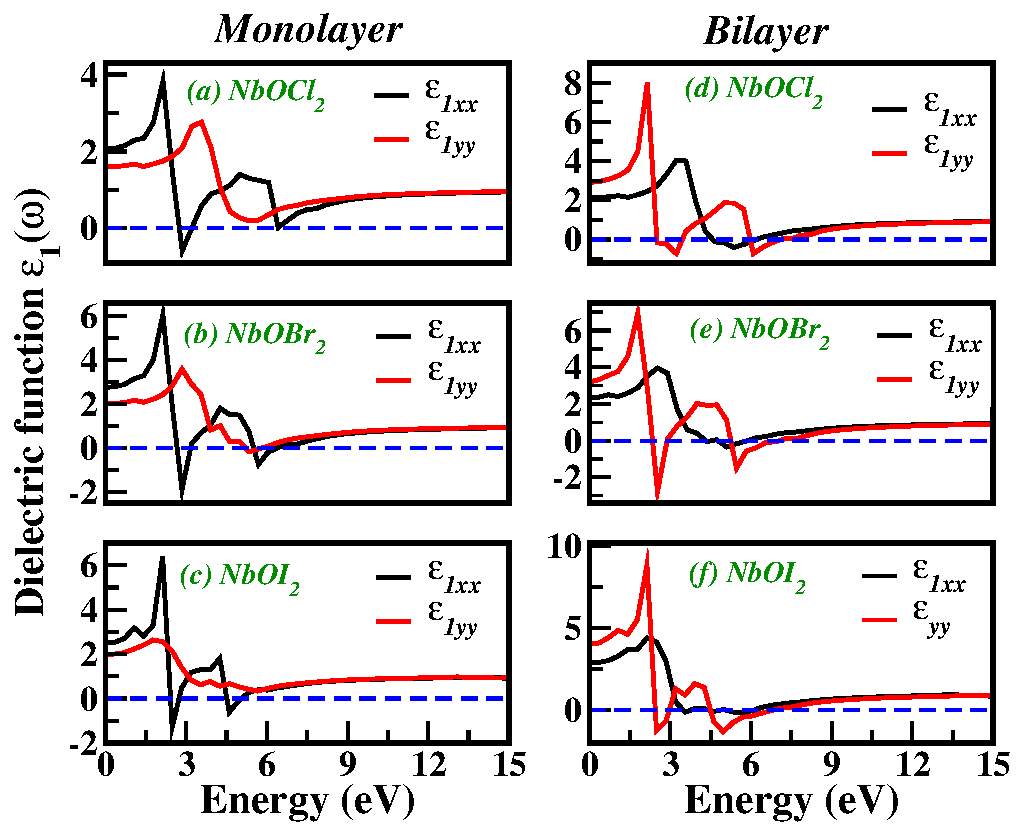}
		\caption{Representation of propagation of electromagnetic waves through the NbOX$_2$ monolayers(a-c) and bilayers(d-f) respectively.}
		\label{7}
	\end{figure}

In this section, we studied the stacking effects on optical properties corresponding to their counterpart monolayers within the PBE-GGA functional. The Fig.\ref{6} and \ref{7} show the dielectric functions along in-plane (xx and yy) directions. It is observed that both the monolayers and bilayers NbOX$_2$ display high anisotropy in the region below 6 eV energy. Whereas, in the higher energy region, the electronic transition along the xx and yy directions is almost similar, which makes them isotropic in nature for both monolayers and bilayers. Such properties were common in 2D materials\cite{du2012hybrid,wang2014enhanced,zhang2019stacking}. 
The peaks in the imaginary part of the dielectric function $\epsilon$$_2$ (in Fig.\ref{6}) are formed due to possible interband electronic transitions from the valence band to the conduction bands. Which is mainly dominated by the halides (X = Cl, Br, and I) p-states in the valence band to the Nb-d states in the conduction band, as observed from PDOS\ref{4}. The dielectric functions(\ref{6} and \ref{7}) have shown the polarisation switching effects similar to other 2D materials\cite{cao2016gate,ji2023general} with stacking. Moreover, optical transition intensities have been increased in the visible and UV region of light, which may be due to more occupied states near the Fermi level. We observed that all the bilayers NbOX$_2$(X = Cl, Br, and I) dielectric function $\epsilon$$_{2xx}$ and $\epsilon$$_{2yy}$ peaks have been enhanced.

\begin{figure}[H]
		\centering
		\includegraphics[width=0.8\textwidth]{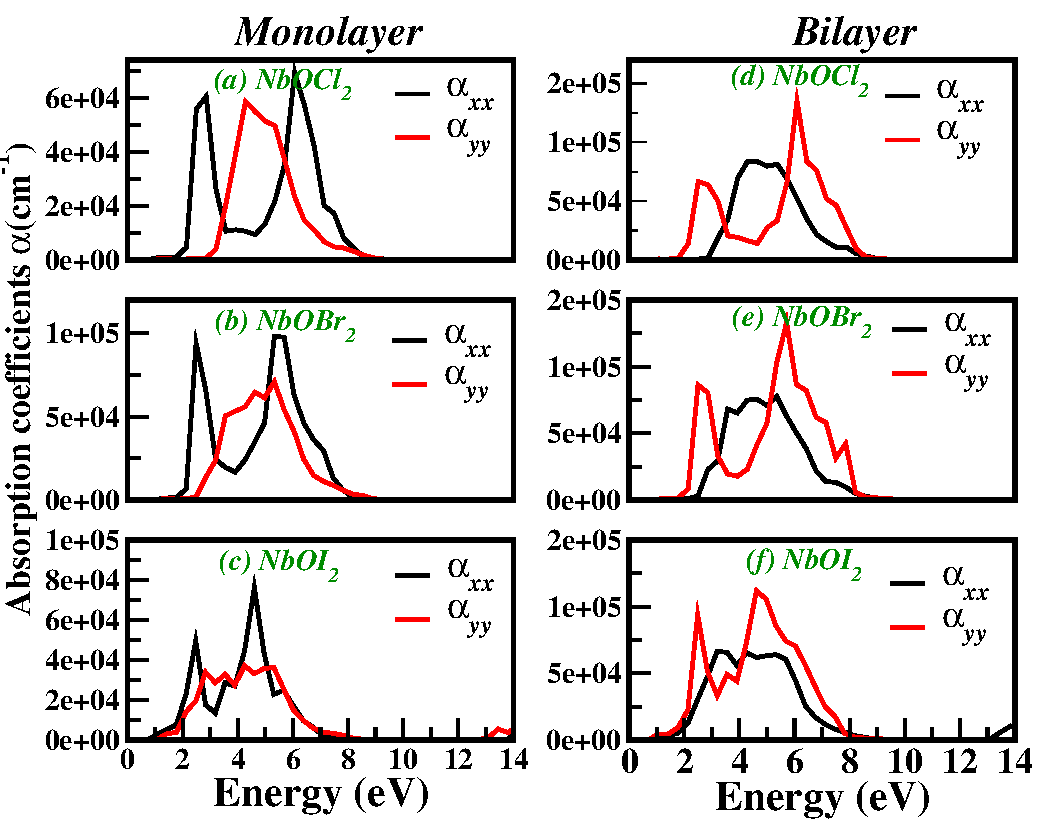}
		\caption{calculated absorption coefficients in the in-plane (xx and yy) directions for NbOX$_2$ monolayers(a-c) and bilayers(d-f) respectively.}
		\label{8}
	\end{figure}
The absorption coefficients of the bilayers have been significantly increased as compared to monolayers(as shown in Fig.\ref{8}) due to interlayer coupling effects, which enhance the optical transition by introducing new electronic states\cite{wu2012visible}.

\begin{figure}[H]
		\centering
		\includegraphics[width=0.8\textwidth]{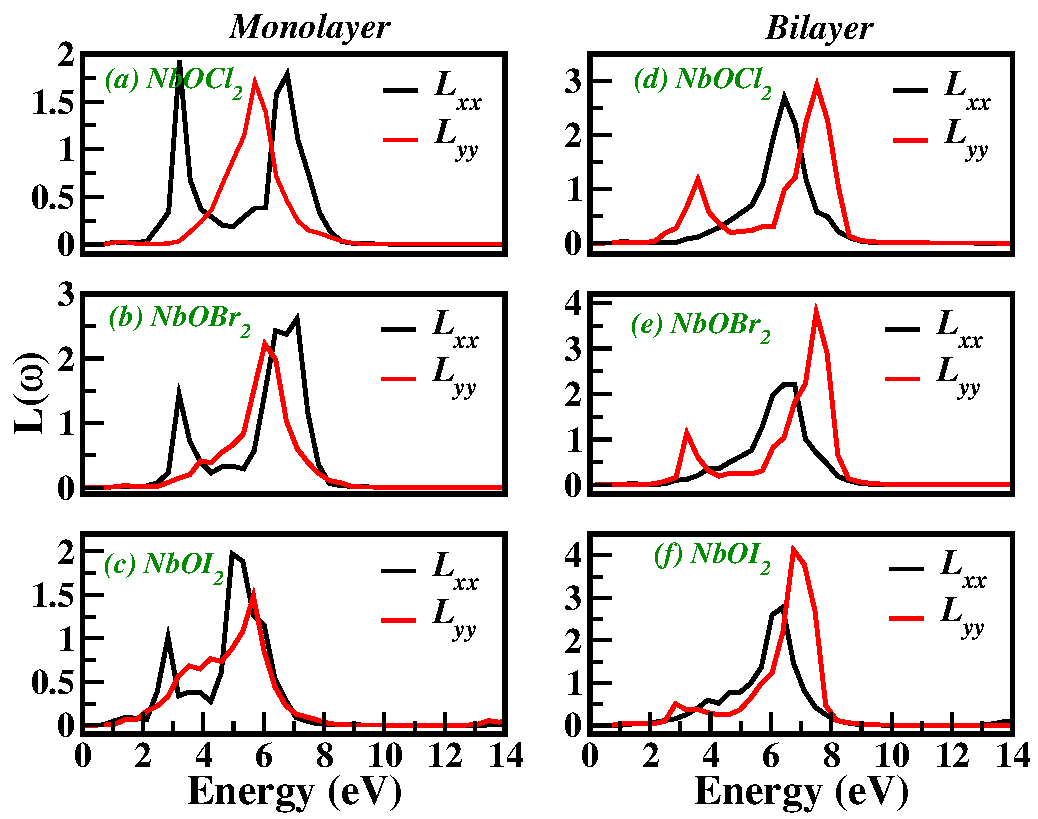}
		\caption{Energy loss function L($\omega$) in the in-plane (xx and yy) directions for NbOX$_2$ monolayers(a-c) and bilayers(d-f), respectively.}
		\label{9}
	\end{figure}
The Fig.\ref{9} depicts the energy loss function given by equation\ref{eq14}, which gives the energy losses by the fast-moving electrons travelling through the medium. Major peaks were observed in the spectra corresponding to plasmon resonances, which arise due to collective oscillations of electrons. In the case of a monolayer, prominent peaks appear in the energy range of (3-8) eV, showing observable anisotropy between L$_{xx}$ and L$_{yy}$ directions. After forming the bilayers, the peak intensities are increased, and a slight increase in plasmon energies is seen. Thus, indicating enhanced interlayer electronic interactions. In bilayers NbOX$_2$(X = Br and I), from the spectral peaks, it is observed that the collective electronic excitations are more pronounced along the y-directions.
It has been observed that homo-bilayer formed under different stacking configurations exhibit enhanced overall optical properties. Thus, we further studied it by using the hybrid functional HSE06, which approximates the experimental value.
The imaginary and the real parts of the dielectric function of homo-bilayer NbOX$_2$ (X = Cl, Br, and I) were evaluated, and the obtained results are represented in Fig.\ref{10}. The absorption onset of Im$\epsilon$ lies within the energy range of 2 to 4 eV, which is in agreement with the previously reported results\cite{mortazavi2022highly}. In these materials, the value of optical absorption is much higher than its band gap energy because the transition is taking place from the inner valence band termed as the extended band gap, avoiding the flat band near the Fermi level to the conduction band \cite{helmer2025computational}. It is also seen that the absorption edge redshifts and peak intensity increase as the halogen atoms get heavier. These effects were observed due to a decrease in electronegativity with increasing p- and d-orbital hybridisation, resulting in a reduced band gap and stronger optical transitions. Moreover, a similar trend was seen in the case of static dielectric constant (Re $\epsilon$), where its value increases with the increase in the atomic number of halogen atoms. The optical absorption coefficients($\alpha$) of homo-bilayer NbOX$_2$ with respect to photon energy are plotted in Fig.\ref{10}. It is seen that the absorption edge of $\alpha$ lies within the visible range of solar energy between (2-4) eV. In addition, these materials exhibit excellent light-harvesting properties within the range of visible to ultraviolet(UV) regions and significant optical absorption coefficients ( $\sim$ 10$^5$ cm$^{-1}$).  

\begin{figure}[H]
		\centering
		\includegraphics[width=1.0\textwidth]{optical.png}
		\caption{schematic representation of (a) Absorption coefficient, (b) Imaginary dielectric function and (c) Real dielectric function of bilayer NbOX$_2$}
		\label{10}
	\end{figure}

\subsection{Transport properties}

After the formation of electron-hole pairs, these photo-generated excitons must reach the redox sites on the material's surface to participate in the overall water-splitting reactions. The mobility of charge carriers must be large enough to promote efficient movement of electron and hole pairs by reducing recombination and enhancing their photocatalytic activity. Since mobility is also an important characteristic for differentiating between ideal catalysts. Therefore, mobility ($\mu_{2D}$) of homo-bilayer NbOX$_2$ was calculated and summarised in Table \ref{t3} using the deformation potential theory introduced by Bardeen and Shockley\cite{bardeen1950deformation}. 
\begin{equation}
    \mu_{2D} = \frac{e\hbar^3C_{2D}}{K_BTm^*m_dE_l^2}
    \label{e4}
\end{equation}
In the above equations \ref{e4}, e, $\hbar$, K$_B$, and T represent the electron's charge, reduced Planck's constant, Boltzmann constant, and room temperature, respectively. m$^*$ is the effective mass of charge carriers given by equation \ref{e5}, which are obtained by fitting parabolic functions to the CBM and VBM of the band structures. It is seen that m$^*$ is inversely proportional to the curvature of its band structure, which means that for a small band curvature, its radius of curvature will be larger, resulting in a large effective mass. On the other hand, for a sharp band curvature, its radius of curvature will be small, leading to a small effective mass. Therefore, the flat bands are always associated with large effective masses, as observed in the Fig.\ref{3} valence band of bilayer NbOX$_2$ were flatter than its conduction band. As a result, the effective mass of m$^*$$_h$ is much larger than the m$^*$$_e$, which corresponds to the highly localized electrons around the Nb atoms near the Fermi level as observed in the Fig.\ref{5}. Thus implying that carriers in the valence bands are subject to stronger correlation effects than those in the conduction band.

\begin{equation}
    m^* = \hbar^2(\frac{\partial^2E_k}{\partial k^2})^{-1}
    \label{e5}
\end{equation}
m$_d$ is the average effective mass{=(m$^*_x$m$^*_y$)$^{1/2}$},
and C$_{2D}$ is the elastic constant given by equation \ref{e6}.
\begin{equation}
    C_{2D} = [{\partial^2E}/{\partial \delta^2}]/S_0
    \label{e6}
\end{equation}
where $\delta$ is the applied uniaxial strain, S$_0$ is the equilibrium area, and E is the total energy of the unit cell. The deformation potential E$_d$ is obtained by linear fitting to the corresponding energy of CBM and VBM with respect to applied strain. 

\begin{table*}[htp]
    \centering

    \begin{tabular}{p{1.35cm} p{1.35cm} c c c c c c}
        
     Systems & Carriers & \multicolumn{3}{|c|}{$\Gamma \rightarrow$ X} & \multicolumn{3}{|c}{$\Gamma \rightarrow$ Y} \\
     \hline
     Bilayer& Charges & m$^*$ & $|E_l|$ & $\mu$ & m$^*$ & $|E_l|$ & $\mu$  \\
     & & (m$_0$) & (eV) & cm$^2$V$^{-1}$s$^{-1}$ & (m$_0$) & (eV) & cm$^2$V$^{-1}$s$^{-1}$ \\
     \hline
     X = Cl & e & 1.206 & 2.78 & 491.51 & 0.326 & 5.49 & 755.52 \\
     & h & 3.576 & 3.61 & 4.31 & -56.946 & 1.72 & 1.93 \\
     
     X = Br & e & 0.932 & 3.61 & 452.93 & 0.318 & 5.15 & 940.30 \\
     & h & -12.234 & 7.02 & 23.88 & -35.35 & 1.64 & 2.18 \\
     
     X = I & e & 0.482 & 5.57 & 489.22 & 0.279 & 5.54 & 1176.9 \\
     & h & -0.974 & 4.70 & 54.95 & 5.286 & 4.73 & 13.77 \\
     \hline
    \end{tabular}
    \caption{Calculated values of effective mass (m$^*$), deformation potential (E), and mobility ($\mu$) of bilayer NbOX$_2$.}
    \label{t3}
\end{table*}
From Table \ref{t3}, it is observed that bilayer NbOX$_2$ shows a slight enhancement in carrier mobility as compared to their respective monolayers due to the interlayer coupling, affecting the electronic properties of the CBM and VBM near the Fermi level. Due to the anisotropic lattice, the electrons are highly mobile along the y-direction, reaching up to 1176.9 cm$^2$V$^{-1}$s$^{-1}$) in NbOI$_2$. Whereas holes are more mobile along the x-direction with the highest value of 54.95 cm$^2$V$^{-1}$s$^{-1}$). This implies that there will be an efficient charge separation, and the maximum number of photoexcited carriers will be available for redox reactions. The calculated mobilities are comparable to those of the other 2D materials\cite{rawat2018comprehensive,zhang2015high,wang2017many,liu2020gen3}.

\subsection{Photocatalytic properties}
\begin{figure}[H]
		\centering 
		\includegraphics[width=0.80\textwidth]{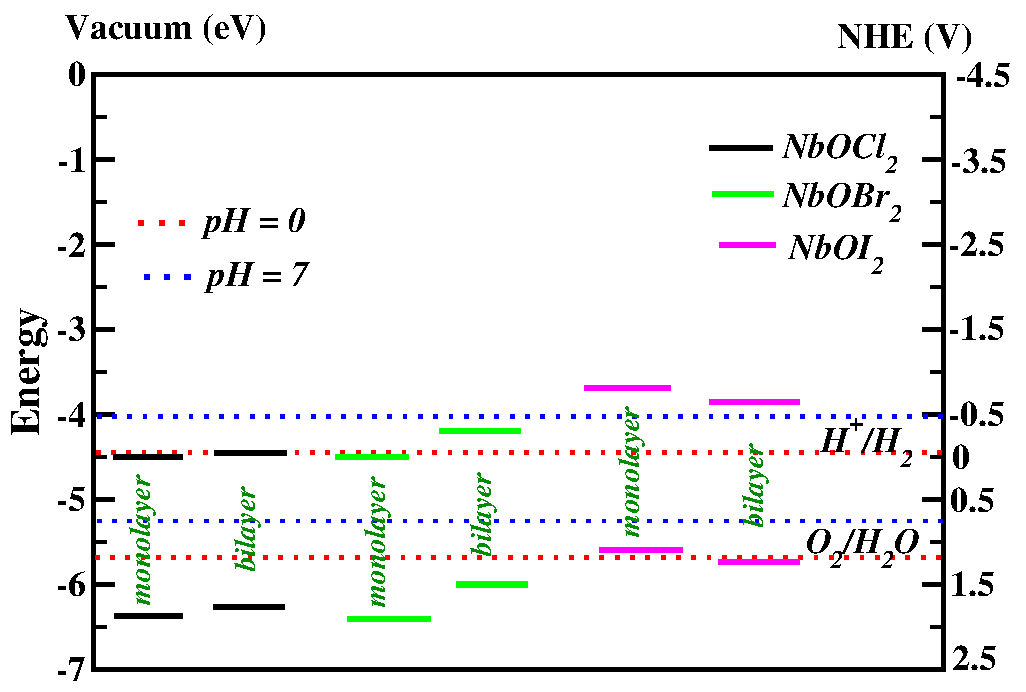}
		\caption{Schematic representation of the band alignment for bilayer and monolayer NbOX$_2$, respectively, at different PH values with respect to the vacuum level.}
		\label{11}
    \end{figure}

The overall photocatalytic splitting of water is a thermodynamically uphill reaction($\Delta$H$_\phi$ = 285.5 KJmol$^{-1}$), which cannot occur spontaneously. It comprises two half-reactions: the oxygen evolution reaction (OER) and the hydrogen evolution reaction (HER). 
For a semiconductor to be an excellent photocatalyst, besides a suitable band gap, it must have a proper band edge potential required to straddle the redox potential of water. To check its band edge potential with respect to vacuum as a reference, we use the equations\cite{xu2000absolute} \ref{a} and \ref{b}.
\begin{equation}
    E_{cb} = - AE = -\chi + 0.5\times E_g
    \label{a}
\end{equation}
\begin{equation}
    E_{vb} = - I = -\chi - 0.5\times E_g
    \label{b}
\end{equation}
where $\chi$ denotes the absolute electronegativity of the semiconductor\cite{butler1978prediction}, $E_g$ refers to the band gap of the semiconductors.
The absolute band-edge positions of VBM (E$_{VBM}$) and CBM (E$_{CBM}$) were obtained using the following equations.\cite{liu2020two,qiao2022two}
\begin{equation}
  E_{VBM}^{abs} = E_{VBM}^{DFT} - E_{vac}  
\end{equation}
\begin{equation}
    E_{CBM}^{abs} = E_{CBM}^{DFT} - E_{vac}
\end{equation}
Where E$_{VBM/CBM}^{DFT}$ are the energies of VBM/CBM obtained by using HSE06. Fig.\ref{11} shows the band-edge potentials of both monolayer and bilayer under investigation. The two red(blue) dashed lines in the figure represent the reduction potential of hydrogen(H$^+$/H$_2$, -4.44 eV) and the oxidation potential of water(H$_2$O/O$_2$, -5.67 eV) at pH = 0, and pH = 7, respectively. One can note that monolayers of NbOCl$_2$ and NbOI$_2$ are not suitable for overall water splitting under acidic conditions due to their misaligned band edges relative to the redox potential. However, under neutral conditions, the NbOI$_2$ monolayer has the required potentials to drive overall water-splitting reactions, consistent with an earlier report\cite{pan2023two}. On the other hand, in the case of bilayer NbOX$_2$, the band edge has been shifted upward towards the more negative side as compared to monolayers.
NbOI$_2$, and NbOBr$_2$, homo-bilayer CBM lies above a hydrogen reduction potential, and the VBM lies below, as compared to the oxidation potential of water relative to the vacuum level, indicating their capability to be used as a photocatalyst for overall water splitting at different environmental conditions. On the other hand, the homo-bilayer NbOCl$_2$ is capable of oxidising water only because its CBM lies below the hydrogen reduction potential relative to the vacuum level.   

To study the OER and HER in detail, we used a computational Standard Hydrogen Electrode (CHE) model proposed by Norskov {\it et al.}\cite{norskov2004origin}, in which the free energy of ($H^+$ + $e^-$ ) is equal to the free energy of ($\frac{1}{2}$H$_2$) molecule was considered at standard condition (i.e., PH = 0, $p$ = 1 bar and T = 298 K) vs standard hydrogen electrode (SHE).

\subsubsection{Oxygen evolution reaction (OER)}
\begin{figure}[H]
		\centering
		\includegraphics[width=0.90\textwidth]{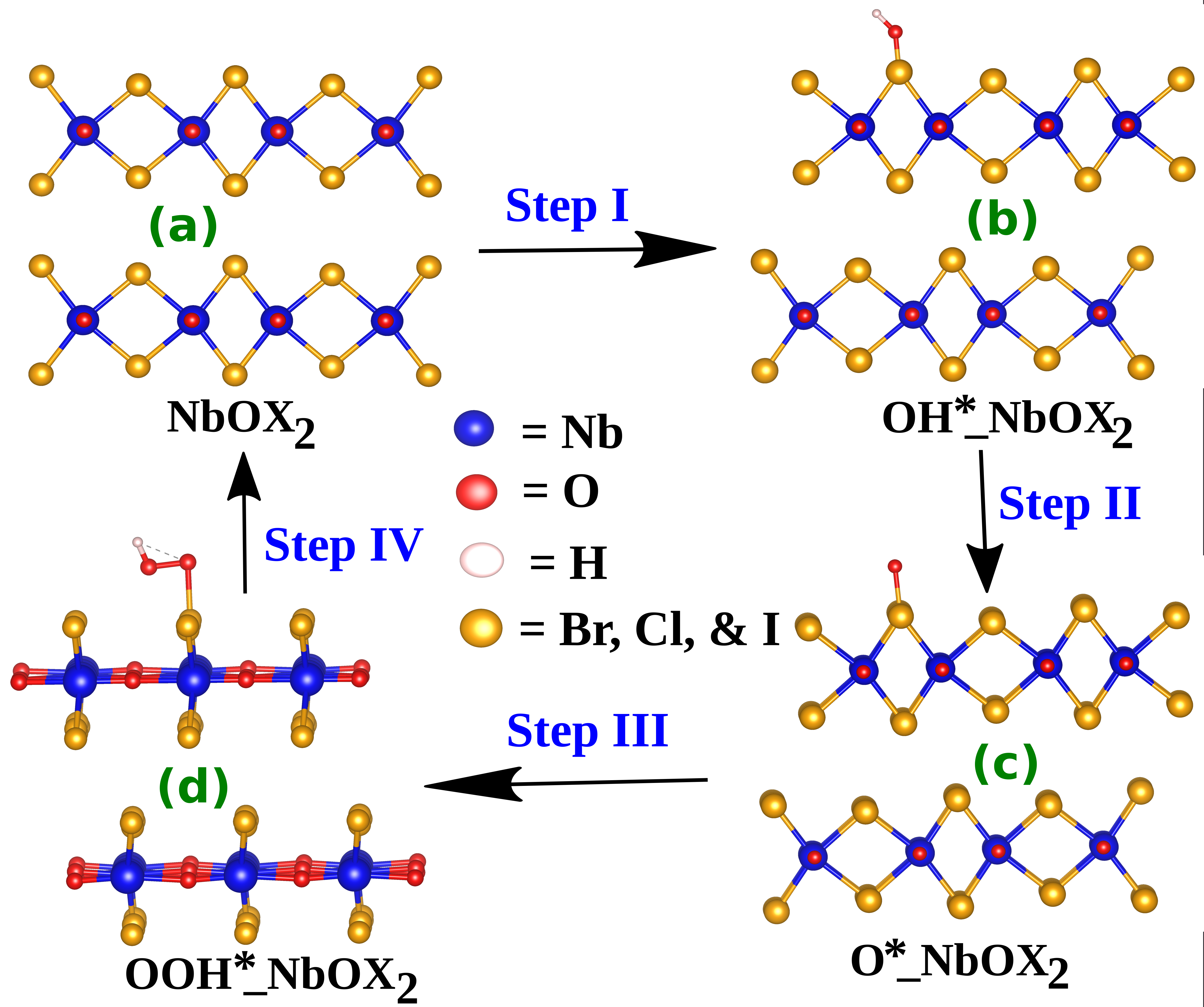}
		\caption{Schematic illustrations of the formation of intermediates in OER.}
		\label{12}
    \end{figure}

During the overall water-splitting process, OER was analysed to be kinetically sluggish due to the involvement of 4 pairs of electrons and protons(as shown in Fig.\ref{12}); therefore, it is considered the rate-limiting step for overall water splitting. The OER takes place at the photoanode electrode of the photo-electrolysis, where photo-generated holes of the valence band oxidise the H$_2$O molecules into the free O$_2$ molecules by the following reaction mechanism \ref{eq31} to \ref{eq34}.\par

i) Adsorption of water molecules on the surface.
\begin{equation}
    H_2O(l) + * \rightarrow OH^* + H^+ + e^-
\end{equation}
\label{eq31}
ii) Oxidation of OH$^*$ species into O$^*$.
\begin{equation}
    OH^*  \rightarrow O^* + H^+ + e^-
\end{equation}
\label{eq32}
iii) O$^*$ combined with its neighbouring water molecules.
\begin{equation}
    O^* + H_2O(l) \rightarrow OOH^* + H^+ + e^-
\end{equation}\label{eq33}
iv) Oxidation of OOH$^*$ species into O$_2$,
\begin{equation}
    OOH^* \rightarrow * + O_2(g) + H^+ + e^-
    \label{eq34}
\end{equation}
where * $\rightarrow$ active sites for adsorption.

To study the stable sites on the surface of the NbOX$_2$ monolayer, we considered six sites (as shown in Fig. S2). From the geometry of the optimised structure after and before adsorption, it was observed that the centre metal atom and oxygen atom sites marked as 1  and 6 (as shown in Figure S3 and Figure S8) overbind the intermediate species leading to distorted geometry, which makes it unstable for the adsorption to occur. The hollow sites marked as 4 and 5 (as shown in Figures S6 and S7) are also not stable during the process, as after optimisation, the adsorbed molecules shift towards the halogen atoms. which indicates that, for monolayer NbOX$_2$ (X = Cl, Br, and I), halogen sites are active for OER (as shown in Figure S4 and Figure S5), consistent with a previous report\cite{pan2023two}. Furthermore, we calculated the free energy profile of OER for the stable active site discussed above.

\begin{figure}[H]
		\centering
		\includegraphics[width=0.70\textwidth]{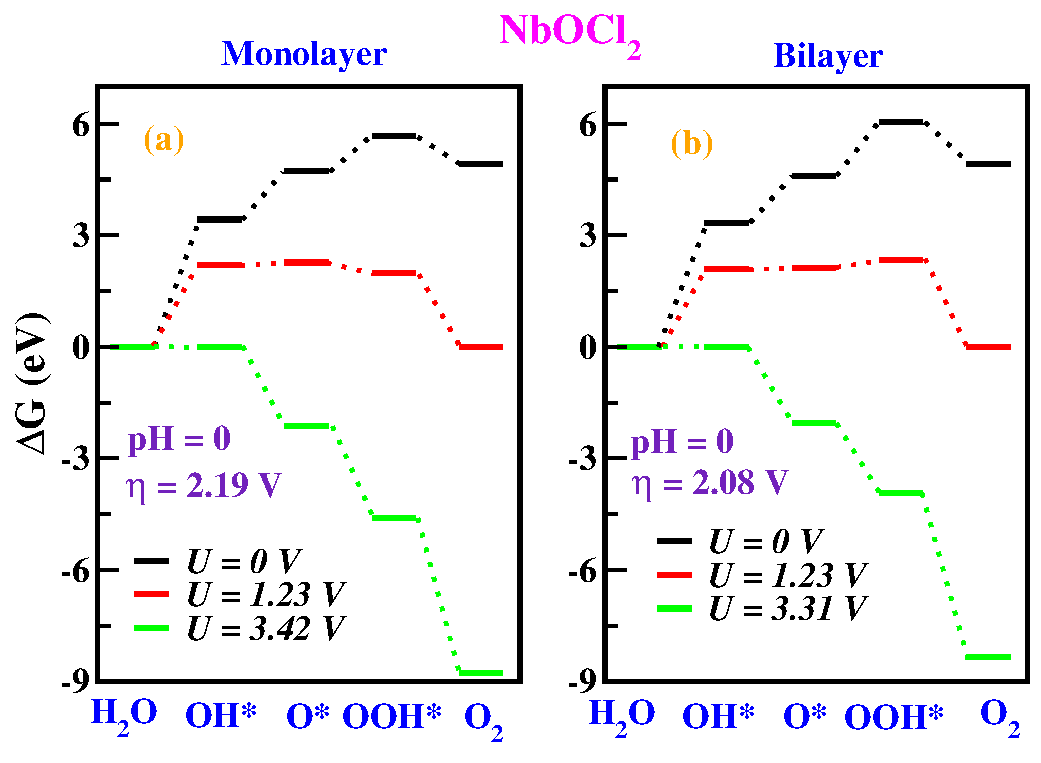}
		\caption{Calculated values of free energy profile of OER: (a) monolayer and (b) bilayer NBOCl$_2$ in acidic conditions.}
		\label{13}
    \end{figure}

\begin{figure}[H]
		\centering
		\includegraphics[width=0.70\textwidth]{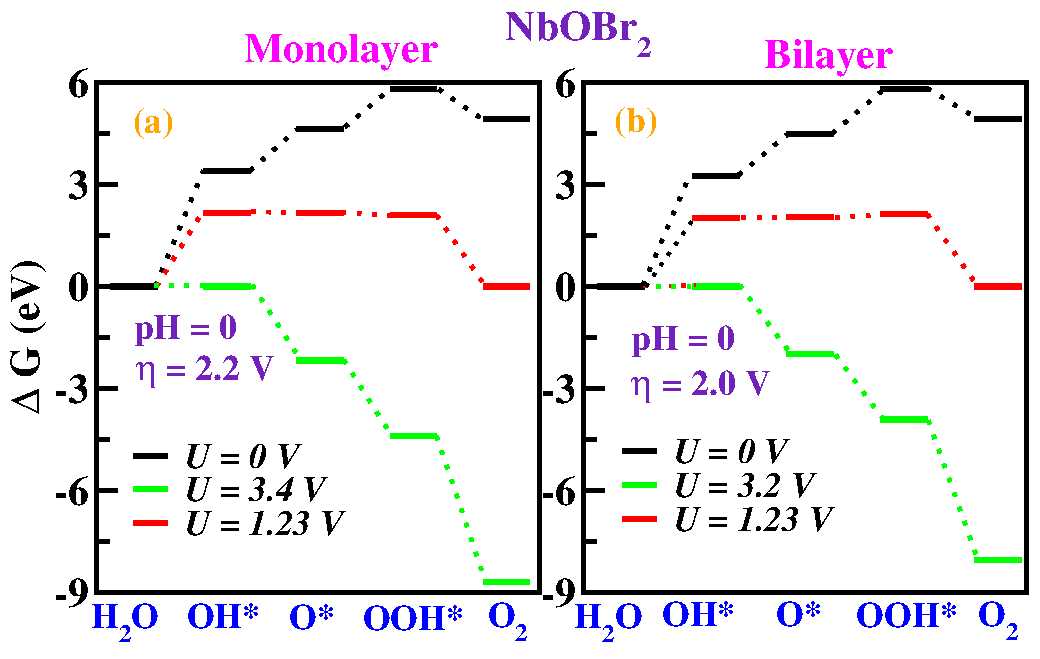}
		\caption{Calculated values of free energy profile of OER: (a) monolayer and (b) bilayer NBOBr$_2$ in acidic conditions. }
		\label{14}
    \end{figure}

\begin{figure}[H]
		\centering
		\includegraphics[width=0.70\textwidth]{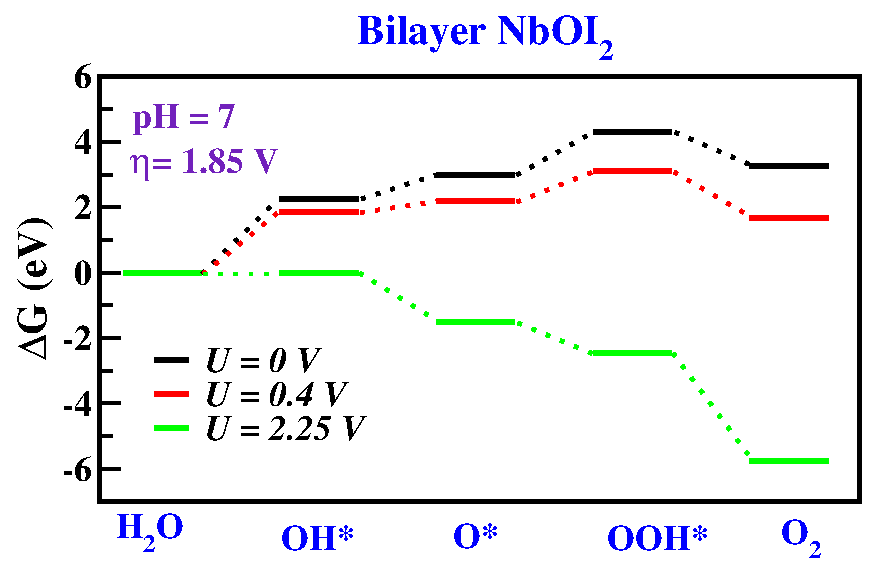}
		\caption{Calculated values of free energy change of OER  for bilayer NBOI$_2$ at PH = 7.}
		\label{15}
    \end{figure}

Fig.\ref{13} gives a plot of the free energy profiles of all the intermediates of OER catalysed by monolayer (\ref{13}a) and bilayer (\ref{13}b) NbOCl$_2$, respectively. It is observed that all reaction steps are endergonic at zero potential. Moreover, even at the equilibrium potential 1.23 V, the first three steps are energetically uphill until the potential increases to 3.42 V for the monolayer and 3.31 V for the bilayer NbOCl$_2$, from which all the intermediate steps became downhill, which means that an extra potential of about 2.19 V for the monolayer and 2.08 V for the bilayer NbOCl$_2$ is needed for OER to proceed spontaneously in acidic medium. The free energy profiles of monolayer (\ref{14}a) as well as bilayer (\ref{14}b) NbOBr$_2$ follow the similar trends as in NbOCl$_2$. Even the potential from which all the intermediate steps become energetically downhill is almost similar, because in pristine metal oxide compounds, the central metal atoms are the most active sites for OER and HER. But in NbOX$_2$ (X = Cl, Br, and I), Nb atoms are structurally shielded by halide atoms, making it difficult to interact directly with the adsorbed molecules. Therefore, the adsorption takes place at the less favourable sites (i.e, halide atoms), which requires a large amount of additional potential beyond the thermodynamic equilibrium potential. But still, these materials show better OER activity as compared to some other materials like S-term MoS$_2$ (2.65 V) and T-term MoS$_2$ (3.39 V)\cite{xu2019design}, V$_p$-BP (3.55 V) and V$_{BP}$-BP (2.49) \cite{zeng2020single} and comparable with (2.06 V) SiP homo-bilayer\cite{somaiya2021potential} and SiAs homo-bilayer (2.06 V)\cite{wang2018electronic}. In bilayer NbOI$_2$ (as shown in Fig.\ref{15}), all the intermediate steps became downhill from U = 2.25 V. Thus, the 1.85 V of external potential is needed for OER to happen spontaneously in neutral conditions. It is also observed that the formation of OOH* is quite difficult in NbOX$_2$ (X = Cl, Br, and I); consequently, the conversion of OH* $\rightarrow$ OOH* became the rate-limiting step for overall reactions. The bilayer NbOX$_2$ (X= Cl, Br, and I) exhibits a lower overpotential as compared to the monolayer, which can be attributed to interlayer stacking effect.    

\subsubsection{Hydrogen evolution reaction(HER)}
The HER takes place on the cathode electrode of the photo-electrolysis, where the rate of reaction depends on the potential of the photoexcited electrons on the conduction band of the electrodes.
These photoexcited electrons reduce the H$_2$O molecules from the electrolyte solution into the H$_2$ molecules, by the following mechanisms\cite{benck2014catalyzing,mir2017comparative}.
1. Volmer reaction: adsorption of H atoms takes place by equations \ref{eq20}.
\begin{equation}
    H^+_{(aq)} + e^- \rightarrow H_{ads)}
    \label{eq20}
\end{equation}
after which is followed by either 1(a) or 1(b) \\
1(a). Volmer-Tafel reactions: adsorbed hydrogen atoms from neighbouring surfaces combine to form the hydrogen molecules given by equations \ref{eq21}.  
\begin{equation}
    H_{(ads)} + H_{(ads)} \rightarrow H_2
    \label{eq21}
\end{equation}
1(b). Volmer-Heyrovsky reactions: hydrogen molecules are formed when adsorbed hydrogen atoms react directly with protons from the solution, given by equations \ref{eq22}.
\begin{equation}
    H_{(ads)} + H^+_{(aqs)} + e^- \rightarrow H_2
    \label{eq22}
\end{equation}
Since both the above mechanisms involve the adsorption of hydrogen atoms $H_{(ads)}$ on the catalyst's surface. Therefore, $H_{(ads)}$ is used to study the catalytic activity of the catalysts\cite{benck2014catalyzing,tsai2015rational}. 
H$_{(ads)}$ energy can be calculated from the adsorption free energy of H($\Delta$G$_H$) from equation \ref{e13}.
\begin{equation}
    \Delta G_H = \Delta E_{(ads)} + \Delta E_{ZPE} - T\Delta S_H
    \label{e13}
\end{equation}

Where $\Delta$E$_{(ads)}$ is the adsorption energy of hydrogen, $\Delta$E$_{ZPE}$ is the zero point energy given by the energy difference between hydrogen in the adsorbed and gas phases, T is the room temperature in K, and $\Delta$S$_H$ is the change in entropy between hydrogen in the gas and adsorbed phases\cite{mir2017comparative}. 

Moreover $\Delta$G$_H$ is used as a descriptor for HER; thus, a ideal catalyst must have $\Delta$G$_H$ $\cong$ 0 eV\cite{tsai2015rational}. Because, if $\Delta$E$_{(ads)}$ is too small, i.e., hydrogen binds weakly to the surface, then HER is limited by Volmer reaction, whereas, if $\Delta$E$_{(ads)}$ is large, meaning hydrogen binds strongly to the surface, which makes the desorption of adsorbed molecules difficult, then HER is limited by Tafel-Heyrovsky reaction. Hence, a good catalyst should not bind 
reaction intermediates either weakly or strongly, fulfilling the Sabatier principle\cite{medford2015sabatier}. The overall reaction mechanism of HER is given below:\\
i) Adsorption of hydrogen atoms on the photocatalyst's surface.
\begin{equation}
    H_2O + e^- \rightarrow H^* + OH^-
\end{equation}
ii) Desorption of Hydrogen molecules.
\begin{equation}
    H^* + H^* \rightarrow H_2
\end{equation}

\begin{figure}[H]
		\centering
		\includegraphics[width=0.76\textwidth]{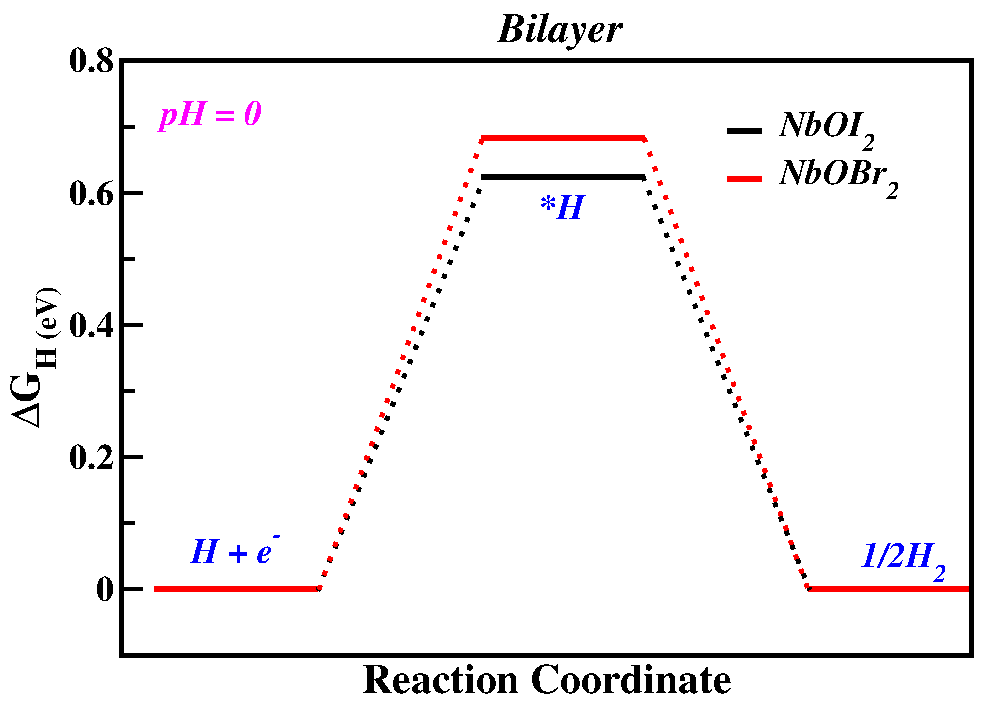}
		\caption{Free energy profile of bilayer NbOBr$_2$ and NbOI$_2$ respectively in acidic environment}
		\label{16}
    \end{figure}

We calculated the hydrogen adsorption energy for monolayer and homo-bilayer NbOX$_2$ using an equation\ref{e16}, which is summarised in Table S4 to check whether our materials under investigation can be used as a photocatalyst for HER or not.
\begin{equation}
    \Delta E_{(ads)} = E_{(s+nH)} - [E_s + \frac{1}{2}H_2]
    \label{e16}
\end{equation}
Where, E$_{(S+n_H)}$ and E$_S$ are the DFT obtained system total energies with and without adsorbed hydrogen, respectively.
Among the considered monolayer and homo-bilayer NbOX$_2$ (X = Cl, Br, and I) compounds under study, 
The homo-bilayer NbOBr$_2$ and NbOI$_2$ fulfilled the criteria for HER. These materials exhibit positive E$_{ads}$ values within the range of (0.44-0.55) eV, implying that hydrogen will form a strong bond with the surface atoms, thus limiting hydrogen production by Tafel/Heyrovsky reactions. Moreover, their free energy has been calculated and shown in Fig.\ref{16}. These values were obtained when oxygen was chosen for an active site for HER, which gave the minimum free energy as compared to other sites required for adsorption and desorption of hydrogen atoms.

\section{Conclusions}
We thoroughly studied the stacking effect on structural, electronic, optical and photocatalytic properties of bilayer NbOX$_2$ (X= Cl, Br, and I). It has been found that bilayer NbOCl$_2$ and NbOBr$_2$ prefer AC stacking, and NbOI$_2$ with AB stacking, exhibits a stable geometric configuration with respect to their ground state energy. Moreover, these bilayers were thermally, dynamically, and mechanically stable. Due to the stacking effect, the band gap of the bilayer under investigation was found to be reduced by a small value without changing its band nature. At the interface of these bilayer NbOX$_2$, moderate charge transfer was observed due to electron redistribution between the layers, which will ultimately help to enhance the photocatalytic efficiency. The bilayer under investigation, specifically NbOI$_2$, shows a higher electron mobility(1176.9 cm$^2$V$^{-1}$s$^{-1}$) along the y-direction and a hole(54.95 cm$^2$V$^{-1}$s$^{-1}$) in the x-direction, leading to reduced recombination centres and more availability of charged species for redox reaction. For bilayer NbOX$_2$ (X = Cl, Br, and I), the overpotential has been reduced by certain amounts. Thus indicating that their overpotential can be reduced by making multilayers similar to BiOI and PtSe$_2$.

\begin{acknowledgement}

\textbf{DPR} acknowledges Anusandhan National Research Foundation (ANRF), a statutory body of the Department of Science \& Technology (DST), Government of India, for the project Sanction Order No.: CRG/2023/000310, dated: 10 October, 2024.  VASP software is under the academic license of Can Tho University, Vietnam.
L.D.Tamang acknowledges the Ministry of Tribal Affairs (Scholarship Division), Government of India, for the support vide No.11019/07/2018-Sch. 
L.D.Tamang also thanks SRM University-AP, Andhra Pradesh, for providing central computational facilities and National Supercomputing Mission (NSM) support for access to PARAM Porul, and for carrying out the computational work.
    

\section*{Author contributions}
\begin{itemize}
\item \textbf{Laku Dorjee Tamang:} First author, Formal analysis, Visualization, Validation, survey of literature, Writing-original draft, writing-review \& editing.
\item \textbf{Shivraj Gurung}:Formal analysis, Visualization, Validation, writing-review \& editing. 
\item \textbf{Bhanu Chettri}:Formal analysis, Visualization, Validation, writing-review \& editing. 
\item \textbf{Nguyen Thanh Tien}:Formal analysis, Visualization, Validation, writing-review \& editing. 
\item \textbf{Le Huu Nghia}:Formal analysis, Visualization, Validation, writing-review \& editing. 
\item \textbf{Darwin Barayang Putungan}:Formal analysis, Visualization, Validation, writing-review \& editing. 
\item \textbf{Ranjit Thapa}: Supervision, Resources, Formal analysis, Visualization, Validation, writing-review \& editing. 
\item \textbf{Kailash Chandra Bhamu}: Supervision, Resources, Formal analysis, Validation, writing-review \& editing. 
\item \textbf{Dibya Prakash Rai:} Project management, Supervision, Resources, Formal analysis, Visualization, Validation, writing-review \& editing. 
\end{itemize}

\section*{Data Availability Statement}
All supporting data for the findings are within the article itself. Further data can be obtained from the corresponding author upon a reasonable request. 

\end{acknowledgement}



\bibliography{achemso-demo}

\end{document}